\documentclass[12pt,a4paper,onecolumn]{revtex4}
\usepackage{graphicx}
\usepackage{amsmath}
\begin{document}

%\catchline{1}{1}{2007}{}{}
%\markboth{Author's Name}{Paper Title}

\title{Electro-optic modulation in a sub-wavelength gap-plasmon guide}

\author{Subimal Deb$^*$, K. Sarat Kumar, P. Anantha Lakshmi, S. Dutta Gupta}

\affiliation{School of Physics, University of Hyderabad, Hyderabad, Andhra Pradesh 500046, India\\
$^*$ subimal.deb@gmail.com
}

\begin{abstract}
We show efficient electro-optic modulation in a subwavelength gap-plasmon waveguide (GPW) formed by an electro-optic polymer with metal coatings. The proposed device is studied in the attenuated total reflection and end-fire configurations. In dealing with the end-fire configuration we used a taper from a micron sized guide to the GPW. The structure is shown to exhibit large phase accumulation over short distances, controllable by the applied modulating voltage. 
\end{abstract}

\maketitle
%\newpage
\section{Introduction}
Transport, confinement and modulation of electromagnetic energy and miniaturization of optoelectronic devices has been the focus of research over the past few years \cite{barnes, atwater2007}. The goal has been to beat the Rayleigh limit and to improve the performance of components like inter/intra chip optical interconnects, low loss waveguides, light emitting devices, modulators and amplifiers for high speed circuits and optical signal processing \cite{hunsperger}. For many of such applications modulation of signals is crucial and requires control over the optical properties, namely, refractive index, of the constituent parts. Thermo-optic, acousto-optic and electro-optic materials have thus been probed for use in intensity and phase modulators \cite{hunsperger, nikolajsen}. Novel materials are being probed in the context of easy integration with electronic circuits and novel chip designs (see for example \cite{ridder, xu2005}). The weak electro-optical properties of silicon have been used in micrometer-sized Mach-Zehnder and ring resonator-based modulators \cite{xu2005, jacobsen2006, barrios}. Electro-optic polymers have been studied in detail \cite{ridder} in micron-sized multilayered structures of poled polymer on glass substrate coated with transparent conducting oxide \cite{park}. Further reduction in the device size requires confinement of electromagnetic energy to sub-wavelength dimensions, posing a challenging problem due to the Rayleigh criterion. It is known that lateral confinement of the modes in sub-wavelength guides using photonic band gap structures or conventional waveguides is not possible \cite{maier}. Metal-dielectric guiding structures supporting surface plasmons and array of metal nano particles, each supporting localized plasmons, have been exploited to this end \cite{ozbay}. The major differences between these geometries have been discussed in detail by Ozbay \cite{ozbay}. A recent work showed the free-space coupling of incident radiation and modulation of reflected light \cite{deng}. The design had a planar geometry of nanometer-sized gold and micron-sized electro-optic polymer layers. Efficient confinement and transportation of light within sub-wavelength dimensions in the microwave and visible regime using arrays of copper rods and gold nanoparticles, respectively, have been demonstrated \cite{maier, maierapl}. An efficient and highly reproducible sub-wavelength scale resonator made of 590nm long ridge on a single crystal Au substrate with surface plasmon polariton modes having a diameter less than 100nm were shown \cite{vesseur}. A numerical study of a 50nm wide silver-cladded air cavity resonator coupled to lateral or inserted cavities capable of functioning as filters for plasmonic waves has been reported \cite{noual}. Efficient propagation in surface plasmon polariton band gap structure of 45nm thick gold film on glass substrate at near-optical wavelengths was shown \cite{bozhevolnyi}. Splitting and combination of line defect modes with the same corrugated gold film surface over 20 $\mu$m long Y-junction was observed. Bend loss analysis indicates their feasibility for photonic circuits. It is now well understood that one of the coupled surface modes in a thin metal film (long range surface plasmon (LRSP)), is associated with larger Q-factor and field enhancement. Some of the recent studies exploit such properties of LRSP to achieve modulation \cite{berini}. A Mach-Zehnder interferometer based structure exploiting LRSP modes was studied at telecommunication wavelengths \cite{nikolajsen}. It consisted of a 15nm thick gold stripe embedded in a thermo-optic polymer and showed efficient modulation and switching properties. LRSP waveguides have been fabricated using a direct wafer bonding method \cite{beriniapl}. The structure consisted of a 20nm thick gold stripe embedded inside a $32\mu$m thick LiNbO$_3$ stack. Electro-optic measurements at a wavelength of 1.55$\mu$m showed that the cladding retains bulk properties even at such small dimensions. Another promising candidate for achieving electro-optic modulation (EOM) in sub-wavelength structures are the nanowire devices. Broadband intensity modulation with high modulation depths were proposed in studies of $\sim$100 nm diameter CdS and GaN nanowires  at optical wavelengths \cite{barrelet, greytak}. Efficient energy confinement and feasibility of short range interconnections with a plasmonic guide were shown both experimentally and by numerical computation \cite{maieradvmat, maierapl, atwater2007}. These studies suggest feasibility of using polymers for efficient modulators due to their ease of integration with several passive materials.
\par
\begin{figure}[h]
\includegraphics[width=5cm]{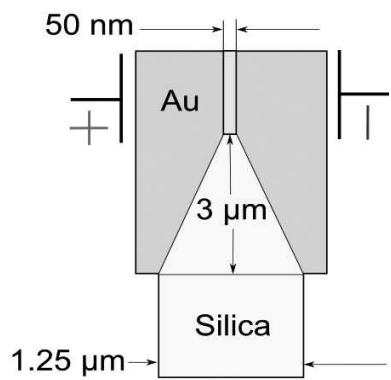}
\caption{  Schematic of the proposed GPW structure.}\label{fig:1}
\end{figure}

 
In this paper we concentrate on a planar geometry of metal-dielectric-metal guides (known as gap-plasmon guides (GPW)) with sub-wavelength lateral dimension in order to show efficient electro-optic modulation. It is known that well confined plasmonic modes in the GPW geometry have longer propagation lengths \cite{bozhebook}. We use the plasmon single-mode regime of such a guide filled with an electro-optic polymer in between the two metal plates. It was shown earlier how the modes of the GPW could be used for slowing down light \cite{sdg}. Here we demonstrate how the same modes can, in principle, be used for achieving EOM in attenuated total reflection (ATR) or end-fire configuration. We show that the device is expected to lead to a large accumulation of phase. Moreover, this phase is shown to be controllable by the applied modulating voltage via the refractive index dependence of the electro-optic polymer on the electric field. The end fire configuration is particularly interesting since in all probability, such is the way a guide is to excited. However, focusing to a spotsize less than 50nm may pose a challenging problem. Besides, the modulator may form a part of a usual communication system. Keeping in view the aforementioned we use a standard 1.25$\mu$m guide (a model for the fiber) followed by a tapered region to the 50nm guide (figure \ref{fig:1}). Recently such a structure was investigated to probe the loss in the micro-to-nano interfacing and a loss of about 30\% was reported \cite{ginzburg}. It is clear that the analytical treatment could be too cumbersome. Hence we resort to numerical finite difference time domain scheme \cite{taflove} (using the commercial product Fullwave \cite{fullwave}) to analyze such a structure. We assume the polymer to be isotropic and the structure to have a zero insertion loss.
\par 
The structure of the paper is as follows. In section \ref{sec:atr} we show how the sensitivity of the surface plasmon resonance on the change of the core refractive index (via the electro-optic effect) can be used for intensity-modulation in reflection in ATR geometry. The next section (\ref{sec:results2}) deals with the FDTD solution for the end-fire configuration. We summarize the important results in Conclusions.

%%%%%%%%%%%%%%%%%%%%%%%%%%%%%%%%%%%%%%%%%%%%%%%%%%%%%%%%%%%%
\section{Electro-optic modulation in the ATR geometry}\label{sec:atr}
\begin{figure}[h]
\includegraphics[width=6cm,height=5cm]{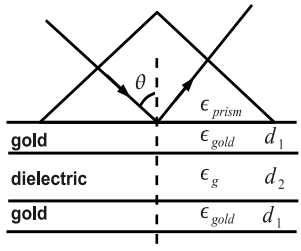}
\caption{  ATR geometry for exciting the gap plasmon structure.}\label{fig:atr}
\end{figure}
It is now well understood that the modes of a planar metal-dielectric structure can be studied in the attenuated total reflection geometry, wherein the guiding structure is loaded on top by a high index prism (figure \ref{fig:atr}). The excitation of the guided and surface modes can be read out in the dips of the reflection coefficient for specific angles of incidence. The location of the dips is extremely sensitive to the guide parameters, the ambient medium refractive index and on the surface conditions. This sensitivity has been exploited to a great deal to realize plasmonic and waveguide sensors and electro-optic modulators (see, for example, \cite{jannsonpat}). In this section we use the specifics of our GPW to show that similar ideas also hold (as expected) at sub-wavelength scales where the guide core width is reduced to about 50nm. Our GPW structure consists of an electro-optic polymer of width $d_2$ sandwiched between two identical gold layers of width $d_1$ (figure \ref{fig:atr}). The polymer is assumed to be isotropic for simplicity and therefore the change in dielectric permittivity ($\epsilon$) can be written as (see, for example, \cite{boyd})
\begin{equation}
\Delta\epsilon = -\epsilon^2 r E  \nonumber
\end{equation}
where $r$ is the electro-optic coefficient, $E$ is the applied electric field. We can further assume a flat response of $\epsilon$ over the range of working frequencies leading to 
\begin{equation}
\epsilon = \epsilon_g (1-\epsilon_g r E) = \epsilon_g (1-E')  \label{eq:1}
\end{equation}
Thus the permittivity of the electro-optic polymer layer has been written down in terms of a dimensionless electric field amplitude ($E' = \epsilon_g r E$) where $\epsilon_g$ is the permittivity of the polymer layer without any applied voltage. 
\begin{figure}
\includegraphics[width=8cm]{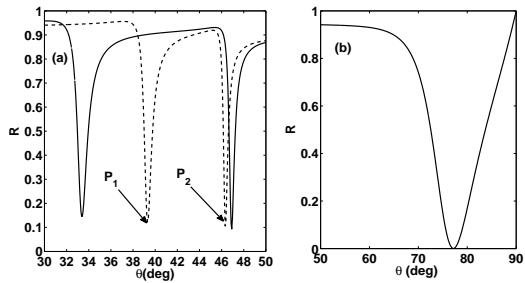}
\caption{ (a) The guided (plasmonic) mode marked P$_1$ (P$_2$) shifts to lower (higher) effective index (i.e., smaller angle of incidence $\theta$) when the $d_2$ is decreased from $0.8\mu m$ (dashed) to $0.6 \mu m$ (solid). (b) Only a plasmonic mode is excited at $d_2=50 nm$ (below the cut-off for the guided modes). All plots correspond to TM polarization. }\label{fig:2}
\end{figure}
\par 
The structure is excited by a monochromatic plane wave incident through a high index prism (refractive index $n_i$) at an angle $\theta$. The modes of such a GPW were studied in detail \cite{sdg, ginzburg}. Below a certain cut-off the only mode that can survive in the structure is the plasmon mode, which can be excited only with p-polarized light. Thus such a guide does not support (at about $d_2=$50nm) any s-polarized waves and essentially is a single mode device below this cut-off width at $\lambda=1.55\mu$m. The calculation of the refraction coefficient for such layered media is fairly standard and we use the characteristic matrix technique to arrive at the magnitude of the intensity reflection coefficient $R$. The results for $R$ as a function of the angle of incidence for applied field $E'=0$ is shown in figure \ref{fig:2} for several values of the guide core thickness (a) 0.8$\mu$m (dashed), 0.6$\mu$m (solid) and (b) 50 nm. Other parameters are as follows: dielectric permittivities of the polymer guide, gold and high-index prism are $\epsilon_g =3$, $\epsilon_\textrm{gold}=-132+12.6i$ and $\epsilon_\textrm{prism}=6.145$, respectively and cladding thickness $d_1=30nm$.
\par
\begin{figure}[h]
\includegraphics[width=12cm]{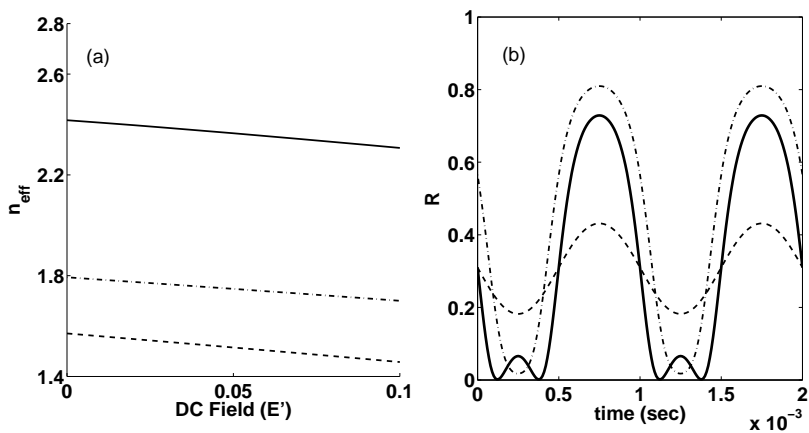}
\caption{ (a) $n_\textrm{eff}$ as a function of the applied field $E'$. The lowest, middle and upper curves corespond to the plasmonic mode at $d_2=0.8\mu m$, $0.6\mu m$ and 50nm, respectively. (b) The AC response at $d_2=50nm$ with $E'=E_0\sin{\omega t}$ at 1kHz. Reflectivity of the GPW structure for $E'=0.05$ at $\theta=$ 73.88$^\circ$ (solid), 72$^\circ$ (dash-dot) and for $E'=$0.01 at $\theta=$ 73.88$^\circ$ (dashed). }\label{fig:acdcresp}
\end{figure}

 
We now study the change in the mode pattern due to the applied voltage and the corresponding change in the reflectivity. It is clear that with a change in the refractive index due to the change in the applied voltage, the resonance dip will get shifted. The effective index ($n_\textrm{eff}$) of a particular mode can be extracted by using the following relation
\begin{equation}
n_\textrm{eff} \sin\theta_\textrm{mode} = n_i \sin\theta_i ~,
\end{equation}
where $\theta_\textrm{mode}$ is the value of the angle of incidence where the dip occurs. The results for the effective index as a function of the normalized dc voltage is shown in figure \ref{fig:acdcresp}(a). It is clear that an increase in the voltage reduces the core refractive index and hence the effective index. A 10\% change in $\epsilon$ caused by the change in the voltage brings about a corresponding change in the effective index.
\par 
The effect of an AC voltage on the intensity reflection can be dramatic, since such a voltage can take the systems to off resonance and again bring it back to resonance. This results in an oscillatory response in the reflectivity (see figure \ref{fig:acdcresp}(b)). The large values of reflectivity correspond to the case where the mode is not excited efficiently, while a low value implies efficient excitation of the guided/plasmon modes. These upper and lower bounds can be regulated by choosing the detuning from the surface surface plasmon resonance. One can thus regulate the efficiency of excitation of the guided/plasmon mode. These features are shown in figure \ref{fig:acdcresp}(b), for $d_2=50nm$ with $E'=E_0\sin{2\pi f t}$ for $f=1$kHz. We show such modulation in figure \ref{fig:acdcresp}(b) over two cycles of the applied AC field. We show this for two amplitudes of the applied field, namely, $E_0=0.01$ and 0.05. At $E_0=0.05$, $\theta=73.88^\circ$ the plasmonic mode is efficiently coupled into the sub-wavelength guide, that is, a good modulation depth is achieved. With $E_0=0.05$, changing $\theta$ to $72^\circ$ (dashed) decreases the modulation depth as is clear from the figure.
%%%%%%%%%%%%%%%%%%%%%%%%%%%%%%%%%%%%%%%%%%%%%%%%%%%
\section{Phase modulation in the end-fire configuration}\label{sec:results2}
\begin{figure}[h]
\includegraphics[width=2.5cm,height=3cm]{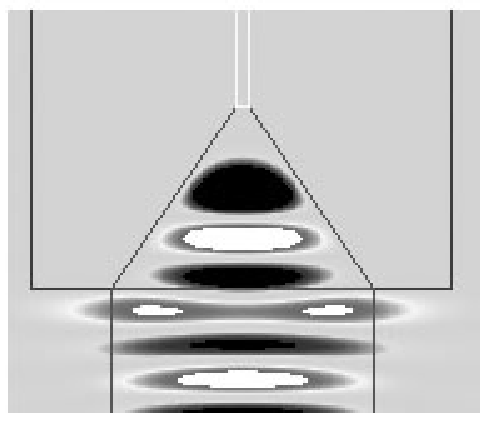}\includegraphics[width=2.5cm,height=2.85cm]{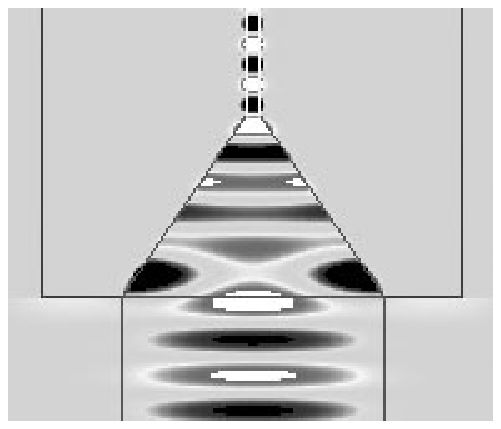}
%\centerline{\psfig{file=fig1, width=5cm}}
\caption{ FDTD simulation of propagation of TE (left) and TM (right) polarized waves through the structure. TE waves are reflected back whereas the TM waves can be coupled into the nano-structure.}\label{fig:fdtd}
\end{figure}
%%%%%%%%%%%%%%%%%%%%%%%%%%%%%%%%%%%%%%%%%%%%%
\begin{figure}[h]
\includegraphics[width=8cm]{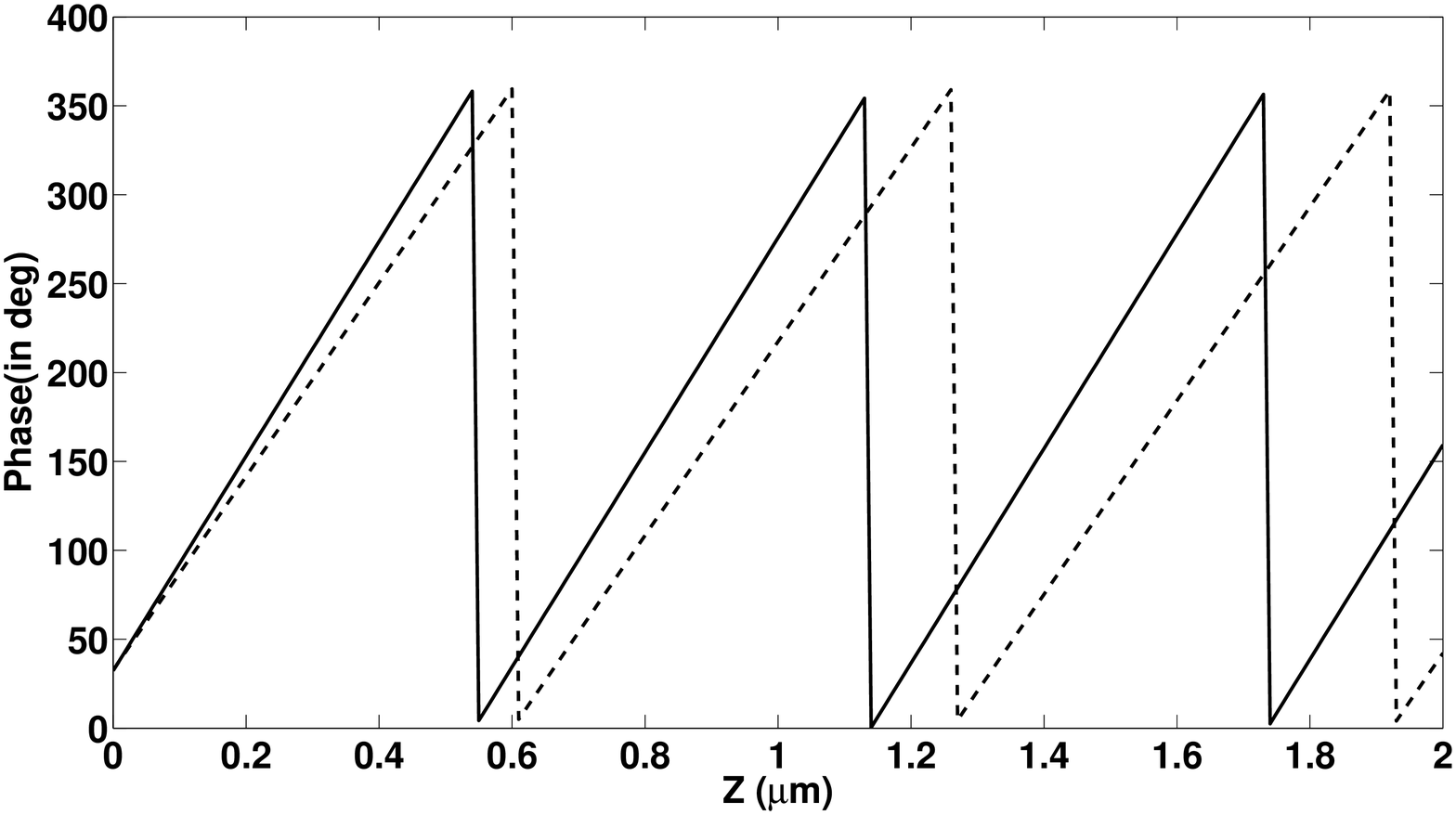}
\caption{ The phase (modulo $2\pi$) of the magnetic field component at the centre of the guide measured along the propagation direction for $E' = $0.1 (dashed) and -0.1 (solid). }\label{fig:phase}
\end{figure}
%%%%%%%%%%%%%%%%%%%%%%%%%%%%%%%%%%%%%%%%%%%%%
Our GPW structure, in practice, need to be interfaced with other components in an optical circuit. It is thus necessary to analyze our structure for the end-fire excitation. We consider coupling of the polymer guide to a standard 1.25$\mu$m silica guide through a taper of height 3$\mu$m. As mentioned earlier, such structures were analyzed in detail for the loss in the tapered region \cite{ginzburg}. It was shown that about 70\% of the incident flux can be coupled into the nano-guide. We study a similar system except that the core is now replaced by the electro-optic polymer. To be specific, we use the FDTD technique to study the coupling of electromagnetic energy from the micro-guide to the 50nm GPW. The simulation again confirms that only the TM polarized modes of the micro-guide can be coupled to the sub-wavelength GPW (figure \ref{fig:fdtd}). The phase accumulated by the plasmon mode propagating along the polymer channel is expected to be proportional to the applied modulating voltage for small variations of $\epsilon$ from $\epsilon_g$ vide equation (\ref{eq:1}). Further, since the electric field in the channel is inversely proportional to the distance between the electrodes, small variations in the applied voltage (amounting to small power consumptions by the circuit) should be sufficient to produce the requisite variations of $\epsilon$ in the sub-wavelength GPW. We extracted the phase of the magnetic field in the plasmon mode as a function of the propagation distance for two voltages corresponding to $E'=\pm 0.1$ in the simulation (see figure \ref{fig:phase}). The calculation of the accumulated phase of the plasmon mode gives a phase difference of about $\pi/2$ for the two values of $E'$ at a propagation length of 1.54$\mu$m. Thus it is shown that, in principle, the GPW configuration can be used effectively as a phase modulator over short distances by controlling the applied field.

\section{Conclusion} \label{sec:conc}
The feasibility of a subwavelength GPW as an efficient electro-optic intensity and phase modulator has been demonstrated by exploiting the single mode regime of the guide, albeit with the assumptions of an isotropic polymer and zero insertion loss. This proposal in principle, demonstrates the possibility of achieving electro-optic modulation at sub-wavelength scales and may find applications in nano-scale integrated optical circuits.


\begin{thebibliography}{99}
\bibitem{barnes} W. L. Barnes, {\it et. al.} {\it Nature} {\bf 424}, 824 (2003).
\bibitem{atwater2007} H. A. Atwater, {\it Sci. Am.} April, p 56 (2007).
\bibitem{hunsperger} R. G. Hunsperger, Integrated Optics, 6th ed. Springer 2009.
\bibitem{nikolajsen} T. Nikolajsen, {\it et. al.} {\it App. Phys. Lett.} {\bf 85}, 5833 (2004).
\bibitem{xu2005} Q. Xu, {\it et. al.} {\it Nature} {\bf 435}, 325 (2005).
\bibitem{ridder} R. M. de Ridder, {\it et. al.} {\it Optical Materials} {\bf 12}, 205 (1999).\bibitem{jacobsen2006} R. S. Jacobsen, {\it et. al.} {\it Nature} {\bf 441}, 199 (2006).
\bibitem{barrios} C. A. Barrios, {\it et. al.} {\it J. App. Phys.} {\bf 96}, 6008 (2004).
\bibitem{park} D. H. Park, {\it et. al.} {\it Optics Express} {\bf 14}, 8866 (2006).
\bibitem{maier} S. A. Maier, {\it et. al.} {\it Nat. Mat.} {\bf 2}, 229 (2003).
\bibitem{ozbay} E. Ozbay {\it Science} {\bf 311}, 189 (2006).
\bibitem{deng} X. Deng {\it J. Opt. A: Pure Appl. Opt.} {\bf 10}, 015305 (2008).
\bibitem{maierapl} S. A. Maier, {\it et. al.} {\it App. Phys. Lett.} {\bf 78}, 16 (2001).
\bibitem{vesseur} E. J. R. Vesseur, {\it et. al.} {\it App. Phys. Lett.} {\bf 92}, 083110 (2008).
\bibitem{noual} A. Noual, {\it et. al.} {\it J. Phys.: Condens. Matter} {\bf 21}, 375301 (2009).
\bibitem{bozhevolnyi} S. I. Bozhevolnyi {\it App. Phys. Lett.} {\bf 79}, 1076 (2001).
\bibitem{berini} P. Berini {\it Advances in Optics and Photonics} {\bf 1}, 484 (2009).
\bibitem{beriniapl} P. Berini, {\it et. al.} {\it App. Phys. Lett.} {\bf 90}, 061108 (2007).
\bibitem{barrelet} C. J. Barrelet, {\it et. al.} {\it Nano Lett.} {\bf 4}, 1981 (2004).
\bibitem{greytak} A. B. Greytak, {\it et. al.} {\it App. Phys. Lett.} {\bf 87}, 151103 (2005).
\bibitem{maieradvmat} S. A. Maier, {\it et. al.} {\it Adv. Mater.} {\bf 13}, 1501 (2001).
\bibitem{bozhebook} S. I. Bozhevolnyi in "Plasmonic nanoguides and circuits", Ed. S. I. Bozhevolnyi,Pan Stanford Publishing 2008.
\bibitem{sdg} S. Dutta Gupta, {\it Pramana- J. Phys.} {\bf 72}, 303 (2009).
\bibitem{ginzburg} P. Ginzburg, {\it et. al.} {\it Opt. Lett.} {\bf 31}, 3288 (2006).
\bibitem{taflove} A. Taflove and S. C. Hagness, Computational Electrodynamics: The Finite-Difference Time-Domain Method, 3rd  ed., Artech House 2005.
\bibitem{fullwave} http://www.rsoftdesign.com/products.php?sub=Component+Design\&itm=FullWAVE
\bibitem{jannsonpat} Jannson {\it et. al.} US Patent No. 5,067,788 (1991).
\bibitem{boyd} R. W. Boyd, Nonlinear Optics, 3rd ed., Academic Press 2009.



%\bibitem{nalwa} H. S. Nalwa, S. Miyata, Nonlinear optics of organic molecules and polymers, CRC Press 1997.

\end{thebibliography}
\end{document}